\documentclass[twocolumn,showpacs,fleqn,nobibnotes]{revtex4}
\usepackage{amsmath,amssymb}
\usepackage{epsfig}
\usepackage{graphicx}

\newcommand{\vecr}{\mathbf{r}}

\newcommand{\vk}{\mathbf{k}}

\newcommand{\ab}{\bar\alpha}

\begin{document}
\title{\bf A simple model for the nuclear unintegrated gluon distribution}
\pacs{12.38.Bx, 13.60.Hb, 12.38.Cy}
\author{M.A. Betemps $^{a}$ and M.V.T. Machado $^{b}$}

\affiliation{
$^a$ Conjunto Agrot\'ecnico Visconde da Gra\c{c}a, CAVG. Universidade Federal de Pelotas, Caixa Postal 460, CEP 96060-290, Pelotas RS, Brazil\\
$^b$ Centro de Ci\^encias Exatas e Tecnol\'ogicas, Universidade Federal do Pampa. Campus de Bag\'e, Rua Carlos Barbosa. CEP 96400-970. Bag\'e, RS, Brazil
}

\begin{abstract}
The solution of non-linear evolution equations for dense nuclear gluon density has been suggested as one of the relevant mechanisms of $pA$ and $AA$ collisions at collider energies. Here we study a simple parameterization for the unintegrated gluon distribution using the knowledge of asymptotic solutions of the Balitsky-Kovchegov equation, describing high-energy QCD in the presence of saturation effects. A satisfactory description of nuclear shadowing at small-$x$ is obtained  and it allows us to understand the qualitative behavior shown by data.
\end{abstract}

\maketitle

\section{Introduction}

The knowledge of the QCD dynamics at high energies is essential in understanding the hadronic interactions studied at current (HERA, Tevatron) and future (LHC) accelerators. In the physics of saturation \cite{glr,muelqiu,ahm90,jkmw97,agl,kovmuel98} the relevant observable is the forward scattering amplitude of a quark-antiquark dipole off a target. It enters several processes at high energy like deep inelastic scattering (DIS), photoproduction and hadron-hadron scattering. Its quantum evolution has recently been the object of many studies \cite{ev1,ev2,ev3,mp}. The Balitsky-Kovchegov (BK) equation \cite{bal,kov} valid in the large $N_c$ limit and in the mean field approximation provides a tool to study the rapidity behavior of the gluon distribution in a large momentum range including the saturation region where the effects of gluon recombination, taken into account in the non-linear term, become important. The main features of analytical solutions of BK have been used to parameterize \cite{marcos} the unintegrated gluon distribution, $\varphi (k,Y)$, which is the Fourier-transform of dipole-target amplitude. The knowledge of the corresponding nuclear unintegrated gluon distribution is indeed crucial. It is the basic object required to calculate physical observables in the $k_{\perp}$-factorized picture employed by the gluon saturation models. In principle, $\varphi (k,Y)$ possesses a Bjorken-$x$ dependence determined (here, $Y=\log(1/x)$) by nonlinear evolution equations of the CGC theory \cite{CGCth} and its dependence on transverse momentum $k$ is fixed by a characteristic saturation momentum, $Q_{s}(Y)$. In most of CGC approaches, the gluon distribution is suppressed below the saturation scale $\varphi_A \propto \log (Q_{s}^2/k^2)$ \cite{KLN} compared to the perturbative form $\varphi_A\propto k^{-2}$. The physics of dense partonic systems and their non-linear perturbative evolution to higher energy has motivated several attempts at understanding bulk properties of ultra-relativistic heavy ion collisions such as the multiplicity, rapidity distribution e centrality dependence of particle production \cite{KLN}. Notice that saturation effects has been claimed to account for the suppression of the high-$p_T$ hadronic spectra in gold-gold and deuterium-gold collisions at RHIC. For the evolved gluon distribution, the yield of produced gluons in $pA$ and $AA$ collisions at central rapidity can be calculated in the factorized way \cite{glr},
\begin{eqnarray}
\frac{dN_{pA}}{dyd^2p_Td^2b} & \propto & \frac{1}{p_T^2}\int d^2k \,\varphi_p (k,y,b)\, \varphi_A (k^{\prime},y,b), \nonumber \\
\frac{dN_{AA}}{dyd^2p_Td^2b} & \propto & \frac{A^{2/3}}{p_T^2}\int \,d^2k \varphi_A (k,y,b) \,\varphi_A (k^{\prime},y,b),\nonumber
\end{eqnarray}
where $k^{\prime}=(k-p_T)$. Under the assumption of local parton-hadron duality, the multiplicity in $AA$ collisions at central rapidity rises proportional to the nuclear saturation scale, $dN_{AA}/d\eta \propto Q_{s,A}^2(\eta=0)$ \cite{KLN}.  For collision energies reaching to 5.5 TeV, the upcoming program in lead-lead collisions at the CERN Large Hadron Collider (LHC), the confirmation of the  conclusions obtained from RHIC is expected to take place. It is also expected the  discrimination between the different physical mechanisms proposed to explain particle production in high energy nuclear reactions. As noticed in Ref. \cite{Javi1}, purely empirical parameterizations of multiplicity data of a large variety of colliding systems allow a logarithmic dependence on collision energy. However, RHIC data by themselves do not differentiate between this and other functional forms like power-laws, negating any possibility to usefully constraint the expectations for LHC energies without further theoretical guidance.

In this work we propose a parameterization for the nuclear gluon distribution based on the numerical/analytical solutions of the BK evolution equation. Starting with a  a proton target, we present a parameterization which encodes the main features of analytical solution of BK equation as geometric scaling property and inclusion of sub-leading terms. It describes all data on DIS in the region of $x<0.01$ in a wide interval of photon virtualities. Relying on geometric scaling arguments, we propose a simple model for the nuclear unintegrated gluon distribution. Such a model is then compared to data on nuclear ratio and its relation to saturation models is investigated. 

\section{A model for the unintegrated gluon function}\label{sec:model}
Lets start by considering the collision between a virtual photon and a proton at high energy. In a frame where the photon travels fast, one can consider that it fluctuates into a $q\bar{q}$ dipole \cite{dipolepic}. The lifetime of this dipole being much longer than the time of interaction with the proton, one can write the cross section as a product of the wavefunction for a photon to go into a dipole times the dipole-proton cross section, which leads to usual formula \cite{dipolepic}
\begin{equation}
\sigma_{T,L}^{\gamma^{*}p}(Q^2,Y)=\int d^2r\int_{0}^{1}dz\,\left|\Psi_{T,L}(r,z;Q^2)\right|^{2}\sigma_{\text{dip}}^{\gamma^{*}p}(r,Y),\label{cross_section}\end{equation}
where $\sigma_{\text{dip}}^{\gamma^{*}p}(r,Y)$ is the dipole-proton cross section. The transverse and longitudinal photon wavefuctions in this expression are computable in perturbative QED.  The proton structure function $F_2$ can be obtained from the $\gamma^*p$ cross section through the relation $F_2=Q^2/(\pi e^2) [\sigma_T+\sigma_L]$.

If one treats the proton as a homogeneous disk of radius $R_p$, the dipole-proton cross section in Eq. (\ref{cross_section}) is usually taken to be proportional to the dipole-proton forward scattering amplitude ${\cal N}(r,Y)$ through the relation $\sigma_{\text{dip}}^{\gamma^{*}p}(r,Y)=2\pi R_{p}^{2}\:{\cal N}(r,Y)$. Here, $Y=\log(1/x)$ is the rapidity variable. The BK evolution equation describes the high-energy evolution of the dipole-proton scattering amplitude ${\cal N}$, where the asymptotic behavior of its solutions is naturally expressed in momentum space. Therefore, the photo-absorption cross section can be written in terms of $\varphi (k,Y)$, the Fourier transform of ${\cal N}(r,Y)$:
\begin{equation}\label{eq:tk}
\varphi (k,Y)=\frac{1}{2\pi} \int \frac{d^2r}{r^2}\,e^{i\vk.\vecr}\,{\cal N}(r,Y).
\end{equation}

Accordingly, the proton structure function can be expressed in momentum space:
\begin{eqnarray}\label{eq:final_f2}
F_{2}(x,Q^{2})=\frac{Q^2}{4\pi^2\alpha_{em}}\int_{0}^{\infty}\frac{dk}{k}\int_{0}^{1}dz|\tilde{\Psi}(k^{2},z;Q^{2})|^{2}\varphi_p(k,Y), \nonumber
\end{eqnarray}
where the wavefunction is now expressed in momentum space as well (see Ref. \cite{marcos} for details). The scattering amplitude $\varphi(k,Y)$, the forward scattering amplitude in momentum space, can be obtained as a numerical result of the BK evolution equation. In case of  neglecting its dependence on the impact parameter, the BK equation can be expressed as
\begin{equation}\label{eq:bk}
\partial_Y \varphi(k,Y) =\ab\chi(-\partial_L)\,\varphi (k,Y)-\bar{\alpha}\,\varphi^{2}(k,Y),
\end{equation}
where $\ab = \alpha_s N_c/\pi$ and $\chi(\gamma)=2\psi(1)-\psi(\gamma)-\psi(1-\gamma)$ is the characteristic function of the Balitsky-Fadin-Kuraev-Lipatov (BFKL) kernel \cite{bfkl} and we used $L=\log(k^2/k_0^2)$ with $k_0$ some fixed soft scale.

In Ref. \cite{marcos}, the properties of analytical solutions of BK equation were used to build an analytical expression for $\varphi (k,Y)$. For instance, the BK equation reduces to the Fisher-Kolmogorov-Petrovsky-Piscounov (F-KPP) equation \cite{fkpp} (after changing of variables) when its kernel  is approximated in the saddle point approximation {\em i.e.} to second order in the derivative $\partial_L$, the so-called diffusive approximation. The F-KPP equation is a well-known equation in non-equilibrium statistical physics, whose dynamics is used to describe many reaction-diffusion systems in the mean-field approximation. This equation has been extensively studied recently and, in particular, it is known to admit traveling waves as asymptotic solutions \cite{mp}. At asymptotic rapidities, the amplitude $\varphi (k,Y)$, instead of depending separately on $k$ and $Y$, depends only on the scaling variable $\tau_p=k^2/Q_s^2$, where we have introduced the saturation scale $Q_s^2(Y)\propto \exp(v_c Y)$, measuring the position of the wavefront.
As discussed in Ref. \cite{marcos}, a more detailed calculation allows also for the extraction of two additional sub-leading corrections, resulting into the following expression for the tail of the scattering amplitude:
\begin{equation}\label{eq:Ttail}
\varphi_p\left(k,\,Y\right) \stackrel{k\gg Q_s}{\approx}
  \left(\tau_p\right)^{-\gamma_c}\log\left(\tau_p\right)
\exp\left[-\frac{\log^2\left(\tau_p\right)}{2\ab\chi''(\gamma_c)\,Y}\right],
\end{equation}
where $\chi^{\prime\prime}$ denotes the second derivative of the BFKL kernel with respect to $\gamma$ and where the saturation scale contains sub-leading terms. The critical parameters $\gamma_c$ and $v_c$ are obtained from the knowledge of the BFKL kernel alone and correspond to the selection of the slowest possible wave, i.e. $v_c=\ab\chi'(\gamma_c)$. For the LO BFKL kernel, one finds $\gamma_c\simeq 0.6275$ and $v_c=4.88\bar{\alpha}$.

Concerning the property of geometric scaling \cite{gscaling}, when one moves along the saturation line, the behavior of the scattering amplitudes remains unchanged. In addition, the sub-leading corrections in \eqref{eq:Ttail} also play an important role as the last term introduces an explicit dependence on the rapidity $Y$ and hence violates geometric scaling. However, this term in negligible when $\frac{\log^2\left(\tau_p\right)}{2\ab\chi''(\gamma_c)Y} < 1$. This means that geometric scaling is obtained for $\log\left(\tau_p\right) \lesssim \sqrt{2\chi''(\gamma_c)\ab Y}$, i.e., in a window which extends $\sqrt{Y}$ above the saturation scale. It is a remarkable property that, at high-energy, the consequences of saturation are observed arbitrarily far in the tail, where $\varphi (k,Y)$ is much smaller than 1. Notice that Eq. \eqref{eq:Ttail} only gives a description of the tail of the wavefront $\varphi (k,Y)\ll 1$ ($k\gg Q_s$). In the infrared domain, one can show that the amplitude behaves like $\varphi_p (k\ll Q_s)\propto c - \log\left(\sqrt{\tau_p}\right)$, with $c$ being a constant.

In Ref. \cite{marcos},  an analytic interpolation between both  asymptotic behavior of the amplitude was proposed. The final expression for the unintegrated gluon function is given by:
\begin{equation}
\label{eq:Tmodel}
\varphi_p\,(\tau_p,\,Y) = \left[\log\left(\frac{\tau_p+1}{\sqrt{\tau_p}}\right)+1\right] \, \left[1-\exp \left(-T_{\text{dil}}\right)\right].
\end{equation}

This equation is our master formula for the further extrapolation to the nuclear case. In expression above, $T_{\text{dil}}$ is an expression which is monotonically decreasing with $L$ and which reproduces (up to the logarithmic factor), the amplitude for geometric scaling \eqref{eq:Ttail}:
\begin{equation}\label{eq:Tdil}
T_{\text{dil}}(\tau_p,Y)= \exp\left[-\gamma_c\log\left(\tau_p\right)-\frac{\log^2(1+\tau_p)-\log^2(2)}{2\bar{\alpha}\chi''(\gamma_c)Y}\right],
\end{equation}
with $Q_s^2(Y) = k_0^2\,e^{\ab v_c Y}$.  The parameters of model have been determined from a fit of all the last HERA measurements of the proton structure function from H1 \cite{h1}, ZEUS \cite{zeus}, with the analysis being restricted to the following kinematic range $ x\leq 0.01$ and $0.045\leq Q^2 \leq 150\: \textrm{GeV${}^2$}$. In our numerical analysis we use the fit including charm \cite{marcos} ($m_q=0.14$ GeV and $m_c=1.3$ GeV), where $v_c=0.807$, $\chi_c''=2.96$, $k_0^2=3.917\times 10^{-3}$ GeV$^2$ and $R_p=4.142$ GeV.

Having presented the unintegrated gluon distribution for the nucleon case, in what follows  we investigate its extension for nuclear targets relying on geometric scaling properties. Based on geometric scaling arguments and on the main features of numerical solutions of BK equation, in Ref. \cite{ASW} it is assumed that both the energy and the nuclear size (or centrality) dependence on the scattering amplitude $N(r,Y;b)$ can be encoded in the saturation scale $Q_{s,h}(Y,b)$ for any hadron $h$ (proton or nucleus). Then the cross section in Eq. (\ref{cross_section}) can be written as
\begin{eqnarray}
\sigma_{T,L}^{\gamma^{*}h}(Q^2,Y)& = &\pi R_{h}^2\int d^2r\int_{0}^{1}dz\,\left|\Psi_{T,L}(r,z;Q^2)\right|^{2}\nonumber \\
& \times &  2 \int d^2\bar{b}\,{\cal N}_h(rQ_{s,h};\,\bar{b}). \label{cross_sectionA}
\end{eqnarray}

As the wavefunction is proportional to $Q^2$ times a function of $r^2Q^2$, the cross section is only a function of $\tau_{h}=Q^2/Q_{s,h}^2$. This geometric scaling was found to describe all lepton-proton data with $x<0.01$. In order to compare protons and different nuclei, in Ref. \cite{ASW} the following assumption about the impact parameter dependence was considered: it can be scaled by the nuclear radius of the hadronic target $h$, $\bar{b}=b/\sqrt{\pi R_h^2}$, with $R_A=(1.12A^{1/3}-0.86A^{-1/3})$ fm. Then, the condition for geometric scaling in lepton nucleus data is 
\begin{eqnarray}
\sigma^{\gamma^*A}_{tot}\left(\frac{Q^2}{Q_{s,A}^2}\right)= \left(\frac{\pi R_A^2}{\pi R_p^2}\right)\times \,\sigma^{\gamma^*p}_{tot}\left(\frac{Q^2}{Q_{s,A}^2}\right).
\end{eqnarray}

For the dependence on atomic number, the saturation scale in the nucleus grows with the ratio of the transverse partons densities to some power $\Delta$, which was taken as a free parameter:
\begin{eqnarray}
Q_{s,A}^2(Y) = \left(\frac{A\,\pi R_p^2}{\pi R_A^2} \right)^{\Delta}\,Q_{s}^2(Y),
\end{eqnarray}
where $Q_{s}(Y)$ is the saturation scale for a proton target. In Ref. \cite{ASW}, it was found $\Delta =1/\delta$, with $\delta = 0.79\pm 0.02$, which translate into a growth of the nuclear saturation scale faster than $A^{1/3}$ for large nuclei.

Now, we are ready to construct a model for the unintegrated nuclear gluon distribution.  Using the geometric scaling arguments, we can rewrite Eq. (\ref{eq:Tmodel}), by replacing $R_p\rightarrow R_A$ and $Q_{s} (Y)\rightarrow Q_{s,A}(Y)$:
\begin{equation}\label{eq:TmodelA}
\varphi_A(k,\,Y) = \left[\log\left(\frac{\tau_A+1}{\sqrt{\tau_A}}\right)+1\right] \, \left\{ 1-\exp \left[- T_{\text{dil}}(\tau_A,Y)\right]\right\},
\end{equation}
where we have defined $\tau_A = (R_A^2/A\, R_p^2)^{\Delta}\, \tau_p$. We think this procedure is suitable once it has been shown in Ref. \cite{ASW} that the data for nuclear structure functions lie on the region where $\tau_A\leq 10$. Therefore, the condition $\log\left(\tau_A\right) \lesssim \sqrt{2\chi''(\gamma_c)\ab Y}\simeq 1.22\sqrt{\log(1/x)}$ is roughly valid and geometric scaling violation should be small. In next section, we investigate the numerical calculation of the nuclear structure function using Eq. (\ref{eq:TmodelA}) and compare them to the experimental measurements of the nuclear ratios.

\begin{figure}[t]
\includegraphics[scale=0.45]{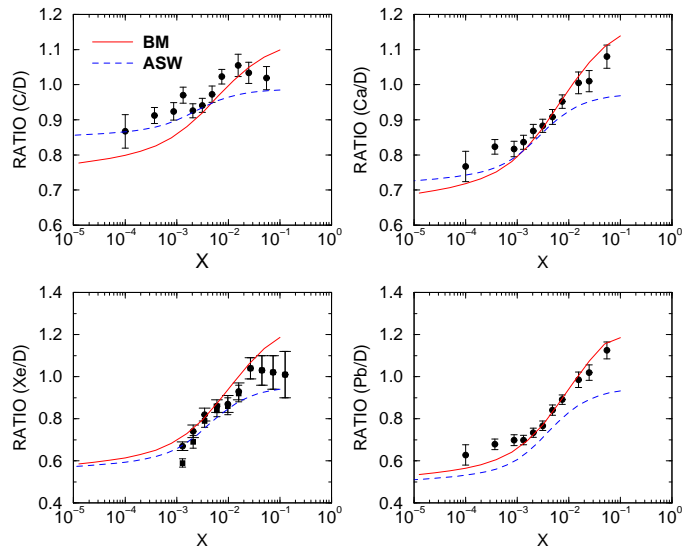}
\caption{The nuclear ratio $R(A/B)=BF_{2}^A/AF_2^B$ as a function of Bjorken-$x$ for several nuclear targets. Solid line represents our results (BM model) and dot-dashed curves the result of ASW model. }\label{fig:1}
\end{figure}

\section{Discussions and conclusions}\label{sec:ccl}
Let us now compare our model for the nuclear unintegrated gluon distribution to the experimental data on nuclear ratios. The shadowing in nuclei is usually studied through the ratios of cross sections per nucleon for different nuclei. In Fig. \ref{fig:1}, the results of the model are compared with experimental data from Refs. \cite{data}. In these figures, the quantity $\mathrm{RATIO}(A/B)=[BF_{2A}(x,Q^2)/AF_{2B}(x,Q^2)]$ is presented. We have selected data where $x\leq 0.01$ and only statistic errors are shown. The solid lines (here labeled as BM model) represents the numerical results using $\Delta = ( 0.79\pm 0.02)^{-1}$. This means that we are using the conservative value of $\Delta$ which reproduces the same growth for the nuclear saturation scale as in Ref. \cite{ASW}.  It is verified that we fairly reproduce the behavior on $x$ for the nuclear ratio, predicting a larger nuclear shadowing for the carbon case. The model is able to describe data even in the region $x\leq 10^{-1}$, taking part of the anti-shadowing region. Considering both the absence of any free parameter in the nuclear unintegrated gluon distribution we find the agreement quite reasonable. For sake of comparison, we show the numerical results using the Armesto-Salgado-Wiedemann (ASW) model \cite{ASW} (dashed lines) for the nuclear cross section. It is verified a saturation of nuclear ration for $x\geq 10^{-2}$ and the model underestimate the data for Lead target. For consistency, we have checked the results for NMC data ($x\geq 0.0125$) and a smaller $\Delta$ values seems to be preferred. This fact has been discussed in Ref. \cite{Raju_Lappi}, where the nuclear enhancement in the saturation scale has been addressed in distinct saturation model (with and without geometric scaling violation). As pointed out in that study, a precise extraction on the $A$ dependence of $Q_{s}$ will play an important role in distinguishing between ``classical'' and ``quantum'' evolution in the CGC. Similar discussion can be found in Ref. \cite{Cazaroto}.

It is timely discuss the qualitative aspects of the nuclear ratios using the present parameterization for the unintegrated gluon distribution. Assuming geometric scaling in the region covered by the experimental data, we can investigate the asymptotic behaviors of the nuclear ratio. At intermediate $x$, around $x=0.1$ the unintegrated gluon distribution has the power-like behavior (modulo logarithmic corrections) $\varphi_A \propto (Q_{s,A}^2/k^2)^{\gamma_c}$ and $\varphi_p \propto (Q_{s}^2/k^2)^{\gamma_c}$ (in fact, for the proton case we have additional scaling violations). This leads to the following result for the nuclear ratio, taking into account geometric scaling arguments:
\begin{eqnarray}
\label{assymp1}
\frac{F_2^A}{AF_2^p}  \propto  \frac{\pi R_A^2}{\pi R_p^2}\,\left[\frac{Q_{s,A}^2(Y)}{Q_s^2(Y)}\right]^{\gamma_c}=  \left(\frac{R_A^2}{A\,R_p^2}\right)^{1-\gamma_c\Delta},
\end{eqnarray}
where the asymptotic behavior seems to be dependent on the product $\gamma_c\Delta$. Notice that in our case $\gamma_c=0.63$ against $\gamma_c=0.75$ for the ASW parameterization. Therefore, the expression above help us to understand the deviations between ASW model and the current model. We have checked that Eq. (\ref{assymp1}) exactly  gives the plateau in $R(A/B)$ seen in the ASW results in the region $x\geq 0.01$. The enhancement observed in our case is a consequence of the anomalous dimension to be dependent on $x$ and $Q^2$ in that kinematic region, where scaling violations start to take place. Therefore, the discussion is more involved in that region, which resembles the discussion on the Cronin peak in $AA$ collisions.

On the other hand, for very small-$x$ (and low $Q^2$), where $k\leq Q_s(Y)$, the unintegrated gluon distribution takes a logarithmic behavior. Thus, one has $\varphi_A \propto \log[(Q_{s,A}^2/k^2)^{\gamma_c}]$ and $\varphi_p \propto \log[(Q_{s}^2/k^2)^{\gamma_c}]$. This leads to the following result for the nuclear ratio,
\begin{eqnarray}
\frac{F_2^A}{AF_2^p} \propto \frac{R_A^2}{R_p^2}\!\left[\frac{\log(Q_{s,A}^2)}{\log(Q_s^2)}\right]=\frac{R_A^2}{A\,R_p^2} \left[ 1+\Delta\frac{\log(\frac{R_A^2}{AR_p^2})}{\log(\frac{Q^2}{Q_{s}^2})}\right]\nonumber \\
\end{eqnarray}
which is clearly dependent on the saturation scale and on photon virtuality. In particular, the ratio seems to be closely independent of anomalous dimension $\gamma_c$. 

As a final discussion, we discuss the relation of present model with recent determinations of the nuclear unintegrated gluon distribution. Notice that the usual notation for the unintegrated gluon density, $f(k,Y)$ is related to function $\varphi (k,Y)$ by the following relation (see for instance Ref. \cite{Kutak_Stasto} for a derivation):
\begin{eqnarray}
f(k,Y)=\frac{N_C}{\alpha_s\,\pi^2}\,\left(1-k^2\frac{d}{dk^2} \right)^2k^2\,\varphi (k,Y).
\end{eqnarray}
There is an comprehensive research program in calculating the numerical solution for the  nuclear gluon distribution from the BK equation including NLO effects such as the running coupling corrections \cite{Javi1,BKnumeric}. From the phenomenological point of view, our  results are similar to the numerical solution as on average the energy behavior for the saturation scale is consistent with $\lambda = 0.2$ and the main features of solutions are present. Recently, a new modified BFKL equation incorporating the shadowing and anti-shadowing corrections of the gluon recombination to the unintegrated gluon function has been proposed through the RSYZ equation \cite{RSYZ}. The numerical results from such an equation is not identical to the BK results in the transverse momentum space but is similar to a mean field result. Of course, our model is less general than gluon function obtained from RSYZ equation as we focus on the shadowing region. Finally, in Ref. \cite{Baier} the authors have tested several parameterizations of unintegrated gluon function in nucleus  inspired on asymptotic solution of BK. Most of them fit our expression for the $\varphi (k,Y)$, which gave satisfactory description of the data for ratio central/peripheral at RHIC.

\section*{Acknowledgments}
This work was supported by the funding agencies CNPq and FAPERGS, Brazil. One of us (MVTM) acknowledges the organizers of the {\it XVII International Workshop on Deep-Inelastic Scattering and Related Subjects -- DIS2009} (Madrid 26-30 April 2009), for their invitation and  hospitality at Universidad Autonoma de Madrid, where  this work has started. We thank Raju Venugopalan for very useful discussions.

\end{document}